\documentstyle[12pt,psfig]{article}

\def\lesssim{\mathrel{\mathpalette\vereq<}}
\def\gtrsim{\mathrel{\mathpalette\vereq>}}
\makeatletter
\def\vereq#1#2{\lower3pt\vbox{\baselineskip1.5pt \lineskip1.5pt
\ialign{$\m@th#1\hfill##\hfil$\crcr#2\crcr\sim\crcr}}}
\makeatother

\begin{document}

\begin{titlepage}
\begin{center}
March 27, 1998     \hfill    LBNL-41631 \\
~{} \hfill UCB-PTH-98/17  \\
~{} \hfill hep-ph/9803481\\

\vskip .1in

{\large \bf Less Minimal Supersymmetric Standard 
Model}\footnote{This work was supported in part by the U.S. Department 
of Energy under Contracts DE-AC03-76SF00098, in part by the National 
Science Foundation under grant PHY-95-14797.  AdG was also supported
by CNPq (Brazil). HM was also supported by the
Alfred P. Sloan Foundation.} 

\vskip 0.3in

Andr\'e de Gouv\^ea,$^{1,2}$ Alexander Friedland,$^{1,2}$ 
and Hitoshi Murayama$^{1,2}$ 

\vskip 0.05in

{\em $^{1}$Theoretical Physics Group\\
     Ernest Orlando Lawrence Berkeley National Laboratory\\
     University of California, Berkeley, California 94720}

\vskip 0.05in

and

\vskip 0.05in

{\em $^{2}$Department of Physics\\
     University of California, Berkeley, California 94720}

\end{center}


\begin{abstract}
  Most of the phenomenological studies of supersymmetry have been
  carried out using the so-called minimal supergravity scenario, where
  one assumes a universal scalar mass, gaugino mass, and trilinear
  coupling at $M_{GUT}$.  Even though this is a useful simplifying
  assumption for phenomenological analyses, it is rather too
  restrictive to accommodate a large variety of phenomenological
  possibilities. It predicts, among other things, that the lightest
  supersymmetric particle (LSP) is an almost pure B-ino, and that the
  $\mu$-parameter is larger than the masses of the $SU(2)_{L}$
  and $U(1)_{Y}$ gauginos.  We extend the minimal supergravity
  framework by introducing one extra parameter: the Fayet--Iliopoulos
  $D$-term for the hypercharge $U(1)$, $D_Y$.  Allowing for this extra
  parameter, we find a much more diverse phenomenology, where the LSP
  is $\tilde{\nu}_{\tau}$, $\tilde{\tau}$ or a neutralino with a large
  higgsino content.  We discuss the relevance of the different
  possibilities to collider signatures.  The same type of extension
  can be done to models with the gauge mediation of supersymmetry
  breaking.  We argue that it is not wise to impose cosmological
  constraints on the parameter space.
\end{abstract}

\end{titlepage}

\newpage

Supersymmetry (SUSY) is regarded as one of the most promising extensions of 
the Standard Model. A supersymmetric version of the Standard Model
will be the subject of exhaustive searches in this and the next generation
of collider experiments.     

The Lagrangian of the minimal supersymmetric extension of the Standard
Model, the so-called ``Minimal Supersymmetric Standard Model'' (MSSM), 
consists of a SUSY-preserving piece and a 
SUSY-breaking piece\cite{MSSM}.  
The SUSY-preserving piece contains
all of the Standard Model parameters plus the so-called $\mu$-term,
once $R$-parity is imposed to prevent baryon/lepton 
number violation.  In this letter, we assume an exact or approximate
$R$-parity, which implies that the lightest supersymmetric 
particle (LSP) does not decay inside detectors.  

The SUSY-breaking Lagrangian will, ultimately, be determined by the 
physics of 
supersymmetry breaking and flavor but at the moment the best approach 
is to simply  parameterize it with a general set of 
explicitly SUSY-breaking parameters. A general explicit soft SUSY-breaking
Lagrangian 
\begin{eqnarray}
 \lefteqn{ \hspace{-0.4cm} {\cal L}_{\rm \begin{picture}(25,0)(0,0)
        \put(0,0){\scriptsize SUSY}
        \put(0,0){\line(4,1){22}}
        \end{picture}}
 = -m^2_{H_d} |H_d|^2 - m^2_{H_u} |H_u|^2 + \left(B\mu H_u H_d+
 \rm{H.c.}\right) } \nonumber \\
&& - \left(
   {\cal A}_d^{ij} \tilde{Q}_i \tilde{d}_j H_d 
  + {\cal A}_u^{ij} \tilde{Q}_i \tilde{u}_j H_u
  + {\cal A}_l^{ij} \tilde{L}_i \tilde{e}_j H_d + \rm{H.c.}\right)\nonumber \\
&&-\sum_{F}m^{2ij}_{\tilde{F}}\tilde{F}^{\dagger}_{i}\tilde{F}_j -
 \sum_{a=1,2,3}(M_{a}\lambda_a\lambda_a + \rm{H.c.}),
  \label{eq:soft}
\end{eqnarray}
where $F=Q,L,U,D,E$ and $i,j=1,2,3$ for each generation, contains
more than 100 new parameters and makes the study of the MSSM parameter
space
completely inviable. Furthermore, a random choice of SUSY-breaking
parameters is most likely ruled out, because of flavor
changing effects and CP-violation. In light of this situation,
simplifying assumptions are not only welcome but necessary. 

The ``minimal supergravity'' framework is the most commonly used set
of assumptions imposed on the MSSM. Because it has 
nothing to do with 
supergravity itself, we will refer to this framework as the
``Very Minimal Supersymmetric Standard Model'' (VMSSM), to avoid confusion.
It assumes a universal scalar mass-squared, gaugino mass, and 
trilinear coupling ($m_{\tilde{F}}^{2ij}=m_{0}^{2}\delta^{ij}$ for all
$\tilde{F}$, $M_a=M_{1/2}$ for all $a$, and ${\cal
  A}^{ij}_f=A_0\lambda^{ij}_f$ for all $f$, where
$\lambda^{ij}_f$ are the ordinary Yukawa couplings) at the grand unified 
(GUT) scale. The VMSSM is, therefore, parameterized by five real
parameters: $m_0^2$, $M_{1/2}$, $A_0$, $\mu$, and $B$ \cite{Betal}.

More recently a lot of work has been done on models with the gauge
mediation of SUSY breaking (GMSB)\cite{GR}. In models of this type again just 
five real parameters are introduced: $F/M$, $M$, $N$, $\mu$, and $B$.
It is important to note that the
particle spectra of models with the GMSB are similar to those of the 
VMSSM \cite{gravitino} and we will, therefore, concentrate our
discussion on the VMSSM and possible modifications to it. 

The issue we would like to address is how restrictive the VMSSM is to
collider phenomenology. 
It is important to be able to explore more diverse particle spectra
while still satisfying all experimental bounds and keeping the
number of parameters small. In this letter we propose a 
``Less Minimal Supersymmetric Standard
Model'' (LMSSM), which adds only one extra parameter to the VMSSM: the
Fayet--Iliopoulos $D$-term for the $U(1)_{Y}$ gauge group, $D_Y$.  Unlike
the VMSSM, this framework will prove to be general enough to allow the
following additional
phenomenological possibilities: a stable charged
slepton, a higgsino-like neutralino, or a sneutrino 
as the LSP. Different particle spectra result in
very different decay patterns, lifetimes and branching ratios which
lead to different signals for SUSY searches, as discussed
later.

A Fayet--Iliopoulos $D$-term for the $U(1)_{Y}$ gauge group is
indeed generated in many interesting theoretical scenarios.
A kinetic mixing between $U(1)_{Y}$ and a different $U(1)$ can
induce a $D$-term once the other $U(1)$ develops a $D$-component
vacuum expectation value\cite{DKM}. The other $U(1)$ can be a part of the
gauge group responsible for dynamical SUSY breaking, or
an anomalous $U(1)$ in superstring theory whose anomaly is canceled
by the Green--Schwarz mechanism. In models with
the GMSB it can also be the messenger $U(1)$\cite{DTW}.
The goal of this letter is, however, to study the effect of the
parameter $D_Y$ on phenomenology, and we will,
therefore, not discuss its origin any further.
 
We will only consider 
constraints from particle
physics. In our opinion it is not wise to impose any 
cosmological constraints on the 
parameter space for the 
experimental analysis of collider data.  Even though cosmology does
provide many useful  
constraints on parameters of particle physics, cosmology at temperatures
between the electroweak scale and nucleosynthesis may be
much more complex 
than usually assumed.  For instance, most models of SUSY 
breaking create cosmological problems, which can be avoided only by 
invoking inflation at temperatures below the electroweak scale\cite{Cosmo}.   
Such a drastic change removes the constraints that the LSP must  
be neutral and should not overclose the Universe.  
Very small $R$-parity violating couplings can also evade the
cosmological constraints without any consequences to collider
phenomenology\cite{R-parity}.
The parameter space should be 
explored without much theoretical prejudice.

\begin{table}
\caption[table]{SUSY-breaking parameters at a scale of 500~GeV
from the 1-loop RG 
equations with the VMSSM boundary conditions at $M_{GUT}=1.86\times 
10^{16}$~GeV, for (A) the first/second 
generation sfermions and (B) the rest with $\tan\beta=10$.  The masses 
of first/second generation fermions have been neglected, and 
$h_t(m_t)=165/(174\sin\beta)$ was used.  The table is to be read as 
follows: each soft parameter is a linear combination of the input 
parameters, with the coefficients given in the table.  For example, 
$m_{H_d}^2 =$ $0.95 m_0^2 + 0.38 (M_{1/2})^2 - 0.01 (A_0)^2 - 0.04 
M_{1/2}A_{0} - 1/2 D_Y$ and $A_{\tilde{d}}=A_0 + 3.41 M_{1/2}$.  }
\label{tbl:RGE}
\begin{center}
\begin{tabular}{|c|c|c|c||c|c|c|} 
\hline
(A)&$m_0^2$&$(M_{1/2})^2$&$D_Y$& & $
A_0$&$M_{1/2}$  \\ \hline
$m_{\tilde{Q}}^2$ &1&5.62&1/6& 
$A_{\tilde{u}}$ & 1 & 3.44\\ \hline
$m_{\tilde{L}}^2$ &1& 0.50&$-1/2$&
$A_{\tilde{d}}$ & 1 & 3.41 \\ \hline
$m_{\tilde{U}}^2$ &1&5.21&$-2/3$&
$A_{\tilde{e}}$ & 1 & 0.67\\ \hline
$m_{\tilde{D}}^2$ &1&5.17&1/3&
- & - & - \\ \hline
$m_{\tilde{E}}^2$ &1&0.15&  1 & 
- & - & - \\ \hline
\end{tabular}
\\ \vskip .1 in
\begin{tabular}{|c|c|c|c|c|c||c|c|c|} 
\hline
(B)&$m_0^2$&$(M_{1/2})^2$&$(A_0)^2$&$M_{1/2}A_0$&$D_Y$& & $
A_0$&$M_{1/2}$  \\ \hline
$m_{\tilde{Q}_3}^2$ &0.63&4.70&$-0.04$&$-0.14$&1/6& 
$A_{\tilde{t}}$ & 0.28 & 2.04 \\ \hline
$m_{\tilde{L}_3}^2$ &$0.99$  & 0.50&$-0.00$&$-0.00$&$-1/2$&
$A_{\tilde{b}}$ & 0.85 & 3.12 \\ \hline
$m_{\tilde{t}}^2$ &0.28&3.45&$-0.07$&$-0.26$&$-2/3$&
$A_{\tilde{\tau}}$ & 0.98 & 0.64 \\ \hline
$m_{\tilde{b}}^2$ &0.97&5.09&$-0.01$&$-0.03$&1/3&
$M_1$& 0 & 0.43 \\ \hline
$m_{\tilde{\tau}}^2$ &0.98&0.14&$-0.01$&$-0.00$&  1 & 
$M_2$& 0 & 0.83 \\ \hline
$m_{H_d}^2$ &$0.95$  & 0.38&$-0.01$&$-0.04$&$-1/2$&
$M_3$& 0 & 2.61 \\ \hline
$m_{H_u}^2$ &$-0.08$  &$-2.15$&$-0.10$&$-0.39$&$1/2$&
- & - & - \\
\hline
\end{tabular}
\end{center}
\end{table}

We first briefly review the VMSSM parameter space and spectrum.
The soft SUSY-breaking
parameters at the weak scale are found by solving the renormalization
group (RG) equations. In Table~\ref{tbl:RGE} we 
quote the results of numerically running the 1-loop RG equations from
the GUT scale down to 500 GeV as a function of 
$m_{0}^{2}$, $M_{1/2}$, and $A_{0}$, for $\tan\beta = 10$ as an
example. The parameters $\mu$ and $B$ run ``by 
themselves'', and one can, therefore, specify their input values at
the weak scale.

It is necessary to check that the electroweak symmetry has been broken
and that $M_Z=91$~GeV. This
is done by choosing $\mu^2$ such that 
\begin{equation}
\label{mu2}
\mu^2=-{M_{Z}^{2}\over 2}+{m_{H_d}^{2}-m_{H_u}^{2}\tan^2\beta\over
  \tan^2\beta-1},
\end{equation}
where $\tan\beta$ is the ratio of Higgs boson vacuum expectation
values, $v_u/v_d$. 
Another condition which must be satisfied 
involves the $B$-term. Once $\tan\beta$ is specified, the $B$-parameter is
uniquely determined and is related to the
pseudoscalar Higgs mass squared, 
\begin{equation}
\label{ma2}
m_A^2=m_{H_d}^2+m_{H_u}^2+2\mu^2=2{B\mu\over \sin(2\beta)}.
\end{equation}
To prevent a runaway behavior in the Higgs scalar potential
$m_A^2$ must be positive. After imposing
Eqs.~(\ref{mu2},\ref{ma2}), the VMSSM contains only four extra real
free parameters: $m_0^2$, $M_{1/2}$, $A_0$, $\tan\beta$, plus a
discrete choice, sign($\mu$).

Table~\ref{tbl:RGE} indicates the structure of the particle spectrum:
colored sparticles are heavier than sparticles
that only transform under $SU(2)_L\times U(1)_Y$ which in turn 
are heavier than
those that only transform under $U(1)_Y$. Furthermore we can numerically
evaluate $\mu^2$ with the help of Eq.~(\ref{mu2}),
\begin{eqnarray}
\label{mu2num}
\mu^2&=&2.18(M_{1/2})^2+0.09m_0^2+0.10(A_0)^2+ \nonumber \\
     &+&0.39M_{1/2}A_0-\frac{1}{2} M_Z^2,
\end{eqnarray}
for $\tan\beta=10$. From gluino searches we find 
$M_{1/2}\gtrsim 77$~GeV
(for $M_3\gtrsim 200$~GeV), and therefore  
$\mu^2\gtrsim 2.14M_2^2$.
We can then safely say that the lightest neutralino 
is an almost pure B-ino of mass $m_{\chi_1^0}\simeq M_1$
\cite{DN}.

There are two LSP candidates: the right-handed scalar tau 
($\tilde{\tau}_R$) and
the lightest neutralino ($\chi_1^0$). 
It is easy to see that $\chi_1^0$ is always the LSP unless
$m_0^2\lesssim (0.04 M_{1/2}^2-1890)$~(GeV)$^2$, for $\tan\beta=10$.
In theories with the GMSB one can
actually have a $\tilde{\tau}_R$ LSP for a larger portion of
the parameter space if the number of messengers ($N$) is large
enough\cite{GR}. 



In the LMSSM the Fayet--Iliopoulos $D$-term ($D_Y$)
changes the mass squared parameters of all the scalars
to $m_{\tilde{F}}^2=m_{\tilde{F},V}^2+Y_{\tilde{F}} D_{Y}$ at some
energy scale, where the
subscript $V$ stands for VMSSM and
$Y_{\tilde{F}}$ is the hypercharge of the scalar $\tilde{F}$. Note
that $Y_{\tilde{F}} D_Y$ is flavor-blind and, therefore, 
the flavor-changing constraints are safely avoided.

There is one very important simplification which is peculiar to the
parameter $D_{Y}$.  $D_{Y}$ runs 
by itself and hence 
it does not matter at what energy scale the scalar masses-squared are
modified.  Therefore, it is convenient to 
calculate $m_{\tilde{F},V}^2$ at the weak scale from the inputs
$m_{0}^2$, $M_{1/2}$, 
and $A_{0}$ (see Table~\ref{tbl:RGE}) and add the weak-scale value of
$Y_{\tilde{F}} D_Y$.  

Similar to the VMSSM, electroweak symmetry breaking imposes
constraints on the parameter space.  One way to 
satisfy Eq.~(\ref{mu2}) is to choose $D_{Y}$ such that 
\begin{equation}
\label{DY}
        \frac{M_{Z}^{2}}{2}+\mu^{2}- 
        \frac{m_{H_{d},V}^{2}-m_{H_{u},V}^{2}\tan^{2}\beta}{\tan^{2}\beta-1}
        =\frac{D_{Y}}{2 \cos2\beta} .
\end{equation}
Note that the form of Eq.~(\ref{ma2}) is unchanged. The free
parameters are, therefore,
\begin{equation}
m_0^2, M_{1/2}, A_0, \tan{\beta}, \mbox{ and } \mu.
\end{equation}
Unlike in the VMSSM, $\mu$ is a free parameter in the LMSSM. It does not,
for example, have to be larger than $M_2$ or even $M_1$. This will
change phenomenology drastically. Note that exactly the same
strategy can be followed to add $D_Y$ as an extra parameter to models 
with the GMSB. 

\begin{figure}[t]
\centerline{\psfig{file=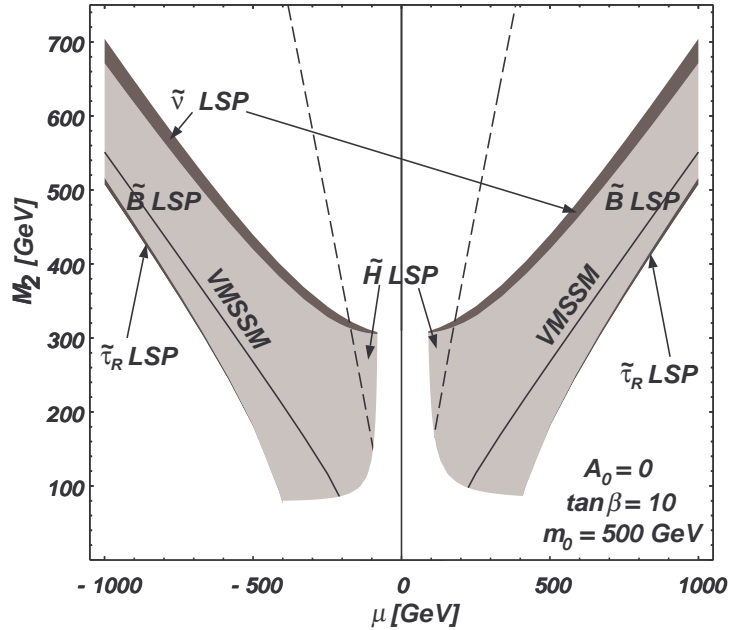,width=0.7\textwidth,angle=0}}
      \caption{Parameter space analysis indicating the nature of the
          LSP. The solid line indicates the points
          allowed by the VMSSM and the dashed line represents points
          where the gaugino content of $\chi_1^0$ is 50\%. $A_0=0$, 
          $m_0^2=500^2$~(GeV)$^2$ 
          and $\tan\beta=10$. The bounds $m_A>65$~GeV, 
          $m_{\tilde{\nu}}>43$~GeV, $m_{\tilde{\tau}}>67$~GeV (if 
          $m_{\tilde{\tau}}<m_{\chi_{1}^{0}}$), and
          $m_{\chi_1^{\pm}}>65$~GeV were imposed.}
\label{fig}
\end{figure}

Varying $D_Y$ (or $\mu$) affects different parameters in
different ways.
For negative $D_Y$, $\tilde{E}$, $\tilde{D}$, and $\tilde{Q}$ become
lighter (the effect on $m^2_{\tilde{D}}$ and $m^2_{\tilde{Q}}$ is,
however, small because of 
their hypercharges), while other sfermions become heavier. 
In this case the absolute value of the $\mu$-term is larger than in 
the VMSSM (see Eq.~(\ref{DY})). If $D_Y$ is large enough
compared to $M_{1/2}$, $\tilde{\tau}_R$ becomes the LSP. Note that,
unlike in the VMSSM, this happens for a large range of values of
$m_0^2$. Fig.~\ref{fig} depicts the
nature of the LSP in the $(\mu,
M_{1/2})$ plane for fixed values of $m_0^2$ and $\tan\beta$. 
For smaller (larger) values of $m_0^2$ or larger (smaller) values of 
$\tan\beta$, the size of the physically
allowed region decreases (increases), but the qualitative features of the 
figure remain 
the same (with the exception of large $\tan\beta \gtrsim 30$, see below). 

For positive $D_Y$, $\tilde{L}$ and $\tilde{U}$ become lighter, 
while all other sfermion masses increase.
In this case the absolute value of $\mu$ is smaller than in
the VMSSM. The consequences of this are many (see Fig.~\ref{fig}).
$\tilde{\nu}_{\tau}$ can become the LSP.  If $\mu$ is
small enough, $\chi_1^0$ can be the LSP but with a large
higgsino content. The mass 
splitting between $\tilde{t}$'s is enhanced with respect to the VMSSM. 
Finally, if
$\tan\beta\gtrsim 30$ and $\mu$ is large, the left-handed $\tilde{\tau}$ can become the
LSP due to left-right mixing in the mass squared matrix.
 
We would like to draw attention to 
the {\it existence} of different particle spectra for different
regions in
the parameter space rather than the size of those regions
 (see Fig.~\ref{fig}). 
Like the VMSSM, the LMSSM should be considered as a parameterization 
and not a model, 
and the fact that diverse spectra can occur is what interests us. 

Next we discuss interesting aspects of the phenomenology
of the spectra outlined above semi-quantitatively.

If $\tilde{\tau}$ is the LSP, heavy stable charged particles become
a good signature for SUSY searches. An analysis of this situation was
done in
the context of models with the GMSB where the $\tilde{\tau}_R$ is the
LSP\cite{FM}. Heavy stable charged particles might
be found by looking for an excess of hits in the muon chambers, or
tracks with anomalously large $dE/dx$ in the tracking chambers.
  
If the LSP is a higgsino-like neutralino, the 
phenomenology is very 
different from the VMSSM case, where the LSP is an almost pure
B-ino. In this case there are four fermions relatively close in mass:
$\chi_1^0$, $\chi_2^0$ and $\chi_1^{\pm}$, which are all 
higgsino-like. In this situation experimental searches are much
harder. Chargino 
searches become more difficult because the mass splitting between
$\chi_1^{\pm}$ and $\chi_1^{0}$ becomes very small
($m_{\chi_1^{\pm}}-m_{\chi_1^{0}}\simeq m_W^2/M_{1/2}$ in the limit of
  $M_2\gg\mu,m_W$), and $\chi_1^{\pm}$ will decay
into missing transverse energy (${\not\!\!E}_T$) plus low energy 
leptons or jets ($E_{l,j}\simeq 6$~GeV if $M_{1/2}=600$~GeV). 
Experimental searches for chargino signals 
at the Tevatron usually require
that $E_T^{l,j}>15$~GeV\cite{Betal,foot}.  

At hadron machines the amount of ${\not\!\!E}_T$ is reduced
because of the small coupling between first and second generation
squarks and $\chi_1^{0,\pm}$.
The main decay mode of a squark is $\tilde{q}\rightarrow
q\chi_{3,4}^{0}$ or $q'\chi_2^{\pm}$, and the heavier
chargino/neutralinos, which are gaugino-like,
further decay via, {\it e.g.}\/, $\chi_2^{\pm}\rightarrow \chi_1^0
H^{\pm}$. The decay chains are
therefore much longer and the amount of ${\not\!\!E}_T$ should decrease. It
is interesting to note that there might be a significant increase in
the number of top quark, b-jet, and $\tau$ events because of the
production of heavy Higgs boson states ($H^{0,\pm}$, $A^{0}$), which have 
large branching ratios into third generation fermions.

The clean tri-lepton
signature at hadron machines will decrease by an order of magnitude
mainly because of the  smaller leptonic branching ratio for $\chi_2^0$ and
$\chi_1^{\pm}$. Note that this effect is
not restricted to the pure higgsino-like neutralino
limit, but also applies to a mixed $\chi_1^0$\cite{Betal}.
 
If the LSP is $\tilde{\nu}_{\tau}$, the decay modes of the heavier
particles change dramatically. There are different possibilities,
depending on $m_{\tilde{l}}$ and $m_{\chi_1^0}$.

If $m_{\tilde{l}}<m_{\chi_1^0}$ the main decay mode for sleptons is
$\tilde{l}\rightarrow \tilde{\nu}jj$ or $\tilde{l}\rightarrow
\tilde{\nu}l'\nu_{l'}$. Charginos, on the other hand, decay into two
particles, namely $\chi^{\pm}\rightarrow \tilde{\nu}l$ or $\rightarrow
\tilde{l}\nu$. The pair production of two sleptons at an $e^+ e^-$
machine will yield, for
instance, $ljj{\not\!\!E}$, which is the typical chargino pair
production signal in the VMSSM. The production of a chargino-pair will yield
acoplanar leptons plus ${\not\!\!E}$, which is the typical slepton
signal at $e^+ e^-$ machines in the VMSSM. 
The two leptons, however, do not have to be of the same
flavor. There are, of course, ways of distinguishing a slepton signature
in the VMSSM from the chargino signal in this scenario because the
cross sections and angular distributions are
quite different.

Another important feature is the visible decay 
$\chi_{1}^{0}\rightarrow \tilde{l} l$.  This makes the production
$q\bar{q} \rightarrow \chi_{1}^{0} \chi_{1}^{0}$ a feasible SUSY 
signature. 
Furthermore squarks decay dominantly as $\tilde{q} \rightarrow q 
\chi_{1}^{0}$ because $\tilde{U}$ is much lighter than $\tilde{Q}$ or $\tilde{D}$,
and hence the squarks produced are
dominantly $\tilde{U}$.  This can 
lead to a rather impressive four leptons plus jets
plus ${\not\!\!E}_T$ signature at hadron machines. 
The total fraction of $4l$ events is only about 0.5\% 
because typically
$BR(\chi_{1}^{0} \rightarrow \tilde{l} l) \simeq 1/3$ and 
$BR(\tilde{\l}\rightarrow \tilde{\nu}_{l} \bar{l}' \nu_{l'}) \simeq 20\%$ for 
$l,l' = e$ or $\mu$, but they have much lower backgrounds \cite{Betal}.

In the case $m_{\tilde{l}}>m_{\chi_1^0}$ both the $\chi_1^{\pm}$ and the
$\tilde{l}$ decay into two on-shell particles ($\tilde{l}\rightarrow
\chi_1^0 l$). The $\chi_1^0$, though unstable, is still invisible,
because its only allowed decay mode is $\chi_1^0\rightarrow
\nu\tilde{\nu}$. This scenario has, therefore, four ``virtual
LSPs'' (3 $\tilde{\nu}$ and the $\chi_1^0$). In this case the
amount of ${\not\!\!E}_T$ in hadron machines is virtually unchanged 
with respect to the VMSSM\cite{Detal}. Note that the clean
tri-lepton signature is enhanced (given that
$\chi_2^0\rightarrow l\tilde{l}$ is allowed with reasonable branching
ratio) because both the $\chi_1^{\pm}$ and the $\tilde{l}$ always decay
into one charged lepton.

Finally, there is another type of signature, which has no VMSSM analog,
if the sneutrino is the LSP and $\tan\beta\gtrsim 4$: visible sneutrino
decays, $\tilde{\nu}_l\rightarrow
\l^{-}\tau^{+}\tilde{\nu}_{\tau}$ . In this case 
the first and second
generation sneutrinos are heavier than $\tilde{\nu}_{\tau}$ enough
to decay visibly. The other allowed sneutrino decays are
$\tilde{\nu}_l\rightarrow
\nu_l\tilde{\nu}_{\tau}\bar{\nu}_{\tau}$ and $\tilde{\nu}_l\rightarrow
\nu_l\tilde{\nu}_{\tau}^*{\nu}_{\tau}$. For $\tan\beta=10$, 
$m_0^2=500^2$(GeV)$^2$,
$m_{\tilde{\nu}_{\tau}}=75$~GeV and $M_1=185$~GeV, $\Delta m\simeq
15$~GeV, and 
the visible branching ratio is approximately $7$\%. In this scenario,
there is a very striking signature for
$\tilde{\nu}_l \tilde{\nu}_l^*$ ($l=\mu,e$) production in $e^{+}e^{-}$
machines if one of the sneutrinos
decays visibly and the other invisibly. One expects to see
$l^{\pm}\tau^{\mp}$ plus ${\not\!\!E}_T$
for $2\times(.07\times.93)=13\%$ of all $\tilde{\nu}_l
\tilde{\nu}_l^*$ produced, for the parameters mentioned earlier. 
The main backgrounds for this signal are $e^+e^-\rightarrow W^+W^-$ and
$\gamma\gamma\rightarrow\tau^+\tau^-$. However, simple kinematic
cuts should efficiently suppress these events, because their 
kinematics are quite different from the signal's. A systematic study of
the appropriate cuts is beyond the scope of this letter. 
There
is also the possibility that $\tilde{\nu}_l$ decays with a displaced
vertex, if $\Delta m$ is small enough. In this case, however, the
visible branching ratio is significantly smaller because of the phase space
reduction due to the tau mass.

In summary, we have shown that the so-called ``Minimal Supergravity
Inspired'' Supersymmetric Standard Model is too restrictive as far as
collider phenomenology is concerned. We proposed the addition of only
one extra parameter to the VMSSM, the Fayet--Iliopoulos $D$-term
for $U(1)_Y$,
and showed that it is capable of yielding a much more diverse
phenomenology while still satisfying all experimental 
constraints.

While the VMSSM almost always yields a B-ino-like LSP, our LMSSM also
allows $\tilde{\nu}$, $\tilde{\tau}$ or Higgsino-like $\chi_1^0$
LSP. We have verified that for each one of these cases there are
important phenomenological consequences, including new signatures for
SUSY and the disappearance of other ``standard'' signatures. 
Even though we do not advocate the LMSSM as {\it the}\/ model of SUSY 
breaking, we emphasize that is a much less restrictive, and yet workable, 
parameterization of the SUSY breaking sector.

\section*{Acknowledgements}
This work was supported in part by the DOE under 
Contracts DE-AC03-76SF00098, in part by the NSF under grant
PHY-95-14797. AdG was also supported by CNPq (Brazil) and 
HM by the  Alfred P. Sloan Foundation.


\begin{thebibliography}{99}
\bibitem{gravitino} The major difference from the VMSSM is that 
 the gravitino ($\tilde{G}$) is most likely the LSP. This can
 lead, {\it e.g.}\/, to photonic signatures from the decay 
 $\chi_1^0\rightarrow \gamma\tilde{G}$.
\end{thebibliography}
\end{document}